\documentclass[11pt]{article}
\usepackage{graphicx}
\usepackage{amsthm,amsmath,natbib}

0.1cm \evensidemargin - 5cm
\newcommand{\eproof}{\rule{0.2cm}{0.2cm}}

\newtheorem{thm}{Theorem}[section]
\newtheorem{prop}[thm]{Proposition}

\newtheorem{cor}[thm]{Corollary}

\newtheorem{remark}[thm]{Remark}

\begin{document}

\title{\Large{{\bf Limit distribution in the $q$-CLT for $q \ge 1$ can not have a compact support}}}
\author{Sabir Umarov$^{1}$
, Constantino Tsallis$^{2,3}$}
\date{}
\maketitle
\begin{center}
$^{1}$ {\it   Department of Mathematics, Tufts University, 503
Boston Ave, Medford, MA 02155, USA\\} $^{2}$ {\it Centro Brasileiro
de Pesquisas Fisicas, and National Institute of Science and Technology for Complex Systems,
Xavier Sigaud 150, 22290-180 Rio de Janeiro-RJ, Brazil}\\
$^{3}$ {\it Santa Fe Institute,
1399 Hyde Park Road, Santa Fe, NM 87501, USA}
\end{center}

\begin{abstract}
In a recent paper Hilhorst \cite{Hilhorst2010} illustrated  that the
$q$-Fourier transform for $q>1$ is not invertible in the space of
density functions. Using an invariance principle he constructed a
family of densities with the same $q$-Fourier transform and claimed
that $q$-Gaussians are not mathematically proved to be attractors.
We show here that none of the distributions constructed in
Hilhorst's counterexamples can be a limit distribution in the
$q$-CLT, except the one whose support covers the whole real axis,
which is precisely the $q$-Gaussian distribution.
\end{abstract}

\vspace{.3cm}



\vspace{1cm}

\renewcommand{\baselinestretch}{2}

\section{Introduction}

Using a specific invariance principle Hilhorst
\cite{Hilhorst2010,Hilhorst2009} showed 
that the $q$-Fourier transform ($q$-FT) is not invertible in the
space of densities. He constructed a family of densities containing
the $q$-Gaussian and with the same $q$-FT. Any density of this
family except the $q$-Gaussian has a compact support.

In the present note we establish that a limit distribution in the
$q$-central limit theorem ($q$-CLT) proved in
\cite{UmarovTsallisSteinberg2008} (see also
\cite{UmarovTsallisGellMannSteinberg2010}) can not have a compact
support. This eliminates all the distributions in Hilhorst's
counterexamples as a valid limiting distribution in the $q$-CLT,
leaving only one, which is the distribution corresponding to the
$q$-Gaussian density. We also show that, for $q>1$, any density which has the
same $q$-FT as the $q$-Gaussian and whose support covers the real
axis is asymptotically equivalent to the $q$-Gaussian.

The $q$-CLT deals with sequences of random variables of the form
\begin{equation}
\label{ZN} Z_N = \frac{S_N-N\mu_q}{\alpha(q)N^{\frac{1}{2(2-q)}}},
\end{equation}
where $S_N=X_1+ \dots +X_N$, {the} random variables $X_1, \dots X_N$
{being} identically distributed and  $q$-independent, 
$\mu_q=\int x [f(x)]^qdx,$ and
$\alpha(q)=[\nu_{2q-1}\sigma^2_{2q-1}]^{1/(2-q)},$ with
\[
\nu_q=\int[f(x)]^q dx, \, \, \sigma_{2q-1}^2=\int (x-\mu_q)^2
[f(x)]^{2q-1}dx.
\]
Without loss of generality we assume that $\mu_q=0$. Three types of
$q$-independence were discussed in paper
\cite{UmarovTsallisSteinberg2008}. Namely, identically distributed
random variables $X_N$ are $q$-independent (see
\cite{UmarovTsallisSteinberg2008}) of\footnote{See
\cite{Tsallis2009a} for definitions and properties of the $q$-product $\otimes_q$
and the $q$-sum $\oplus_q$.}
\begin{align} \label{q-ind1}
\mbox{Type I:} ~~~~
&F_q[X_1+\dots+X_N](\xi)=F_q[X_1](\xi)\otimes_q\dots\otimes_q
F_q[X_N](\xi);\\ \label{q-ind2} \mbox{Type II:} ~~~~
&F_{q}[X_1+\dots+X_N](\xi)=F_q[X_1](\xi)\otimes_{q_1}\dots\otimes_{q_1}
F_q[X_N](\xi); \\ \label{q-ind3} \mbox{Type III:} ~~~~
&F_q[X_1+\dots+X_N](\xi)=F_{q_1}[X_1](\xi)\otimes_{q_1}\dots\otimes_{q_1}
F_{q_1}[X_N](\xi),
\end{align}
if these relationships hold for all $N\ge 2$ and $\xi \in
(-\infty,\infty)$; $q_1=\frac{1+q}{3-q}$.
Here the operator $F_q$ is the $q$-FT defined
as
\begin{equation}
\label{Fq} F_q[X_1](\xi) = \tilde{f}_{q}(\xi):= \int_{-\infty}^\infty
\frac{f(x)\,dx}{[1+i(1-q)x\xi f^{q-1}(x)]^{\frac{1}{q-1}}} \,,
\end{equation}
{with $q>1$}.

We recall some facts about the {$q$-exponential} and
{$q$-logarithmic} functions.
These functions are respectively defined as (see for instance \cite{UmarovTsallisSteinberg2008})
\begin{align*}
\exp_q(x)&=[1+(1-q)x]_+^{{1}\over {1-q}} \;\;\;(\exp_1(x)=\exp(x)) \,,\\
\end{align*}
{and}
\begin{align*}
\ln_q (x)&= \frac{x^{1-q}-1}{1-q}, \, x>0 \;\;\;(\ln_1(x)=\ln(x)).
\end{align*}
It is easy to see (see  \cite{UmarovTsallisSteinberg2008}) that for the $q$-exponential, the relations
$\exp_q(x \oplus_q y) = \exp_q(x) \exp_q(y)$ and
$\exp_q{(x+y)}=\exp_q(x) \otimes_q \exp_q(y)$ hold. In terms of
$q$-log these relations can be equivalently rewritten as follows:
$\ln_q (x \otimes_q y)=\ln_q x + \ln_q y$, and $\ln_q (x y)=\ln_q x
\oplus_q \ln_q y.$ It is not hard to verify that if $1<q_1<q_2,$
then
\begin{align} \label{ln01}
&\ln_{q_{_{2}}}(x) \le  \ln_{q_{_1}}(x) ~ ~ \mbox{for\, all} ~ x>1,
\\
\label{ln02} &\ln_{q_{_{2}}}(x) \ge \frac{q_1-1}{q_2-1}
\ln_{q_{_1}}(x) ~ ~ \mbox{for\, all} ~ x>0.
\end{align}
For $q > 1$ the $q$-exponential is defined for all $x<\frac{1}{q-1}$
and blows up at the point $x=\frac{1}{q-1}.$ The $q$-exponential can
also be extended to the complex plane and it is bounded on the
imaginary axis: $|\exp_q{(iy)}| \leq 1.$ Moreover, $|\exp_q{(iy)}|
\rightarrow 0$ if $|y| \rightarrow \infty.$ Using the
$q$-exponential function, the $q$-FT of $f$ can be represented in
the form
\begin{equation}
\label{identity2} \tilde{f}_q(\xi) = \int_{-\infty}^{\infty}f(x)\,
\exp_q{(ix \xi [f(x)]^{q-1})}\,dx.
\end{equation}
We refer the reader to papers
\cite{UmarovTsallisSteinberg2008,UmarovTsallisGellMannSteinberg2010,TsallisQueiros2007,QueirosTsallis2007,UmarovTsallis2007,UmarovQueiros2010,NelsonUmarov2010,JaureguiTsallis2010a,ChevreuilPlastinoVignat2010,JaureguiTsallis2010b}
for various properties of the $q$-FT.

\section{On the support of a limit distribution}
For the sake of simplicity we consider a 
continuous and symmetric about zero density function $f$ of a random variable $X_1.$ Other
cases can be considered in a similar manner with appropriate care.
Denote $\lambda(x)=x[f(x)]^{q-1},$ where $1 \le q<2.$ Since $f$ is
symmetric, it suffices to consider $\lambda(x)$ only for positive
$x.$ Suppose $\lambda$ attains its maximum value $m$ at a point
$x_m>0,$ i.e. $m=\max_{0<x \le a}\{x[f(x)]^{q-1}\}=x_m
[f(x_m)]^{q-1}. $
\begin{prop}\label{Prop:PW}
Let $f$ be a continuous symmetric density function with $supp \, f
\subseteq [-a,a].$ Then the $q$-FT of $f$ satisfies the following
estimate
\begin{equation} \label{prop1:est}
|\tilde{f}_q(\eta-i \tau)| \le \exp_q(x_m M_q \tau),
\end{equation}
where $\eta \in (-\infty,\infty),$ $\tau < \frac{1}{m(q-1)},$ $M_q=
\max_{[0,a]}\{[f(x)]^{q-1}\},$ and $x_m$ is the point where
$xf^{q-1}$ attains its maximum $m.$
\end{prop}

\textit{Proof.} For $f$ with $supp \, f \subseteq [-a,a],$ equation
(\ref{Fq}) takes the form
\begin{equation}
\label{Fq10} \tilde{f}_{q}(\xi)= \int_{-a}^a
\frac{f(x)dx}{[1+i(1-q)x\xi f^{q-1}(x)]^{\frac{1}{q-1}}}.
\end{equation}
Let $\xi=\eta+i\tau$ where $\eta=\Re(\xi)$ is the real part of $\xi$
and $\tau=\Im(\xi)$ is its imaginary part. We assume that $\eta \in
(-\infty,\infty)$ and $\tau > - \frac{1}{m(q-1)}.$ Then for the
denominator of the integrand in (\ref{Fq10}) one has
\begin{align} \label{denom}
[1+i(1-q)x(\eta &-i\tau) f^{q-1}(x)]^{\frac{1}{q-1}}=[1+i(1-q)\eta
f^{q-1}(x)+(1-q)\tau x f^{q-1}(x)]^{\frac{1}{q-1}} \notag \\&
=[1+(1-q)\tau x f^{q-1}(x)]^{\frac{1}{q-1}}[1+i\frac{(1-q)\eta
f^{q-1}(x)}{1-(1-q)\tau x f^{q-1}(x)}]^{\frac{1}{q-1}} \notag\\
&= \left(\exp_q(\tau x f^{q-1}(x))\right)^{-1}\left(\exp_q(
i\frac{(1-q)\eta f^{q-1}(x)}{1-(1-q)\tau x f^{q-1}(x)}
)\right)^{-1}.
\end{align}
Using the inequality $|\exp(iy)| \le 1$ valid for all $y \in
(-\infty,\infty)$ if $q>1,$ it follows from \eqref{denom} that
\begin{align} \label{denom10}
\Big|1+i(1-q)x(\eta &+i\tau) f^{q-1}(x)\Big|^{\frac{1}{q-1}} \ge
\left(\exp_q(\tau x f^{q-1}(x))\right)^{-1}.
\end{align}
Now, \eqref{Fq10} together with \eqref{denom10} and $f(x)$ being a
density function, yield \eqref{prop1:est}. \eproof

\begin{remark}
Proposition \eqref{Prop:PW} can be viewed as a generalization of the
well known Paley-Wiener theorem. Indeed, if $q=1$ then
\eqref{prop1:est} takes the form
\begin{equation} \label{PW}
|\tilde{f}(\eta-i\tau)| \le \exp(a \tau), \, \eta+i\tau \in
\mathcal{C},
\end{equation}
which represents the Paley-Wiener theorem for continuous density
functions.
\end{remark}
Inequality \eqref{PW} can be used for estimation of the size of the
support of $f.$ Consider an example with
$f(x)=(2a)^{-1}{\mathcal{I}}_{[-a,a]}(x),$ where
$\mathcal{I}_{[-a,a]}(x)$ is the indicator function of the interval
$[-a,a].$ The Fourier transform of this function is
$\tilde{f}(\xi)=(a\xi)^{-1}\sin (a\xi),$ $M_q=M_1=1,$ and $x_m=a.$
Therefore, we have $|\tilde{f}(-i\tau)| \le e^{a\tau}, \tau >0.$
This inequality yields $$2a \ge 2 \sup
\frac{\ln|\tilde{f}(-i\tau)|}{\tau},$$ which gives an estimate from
below for the size $d(f)=2a$ of the support of $f.$

This idea can be used to estimate the size 
of the support of $f$ using the $q$-FT and Proposition
\ref{Prop:PW}. Namely, inequality \eqref{prop1:est} with $\eta=0$
implies
\begin{equation} \label{suppsize}
d(f)=2a \ge 2x_m \ge \frac{2}{M_q} \sup_{\tau} \frac{\ln_q
|\tilde{f}_q(-i\tau)|}{\tau}.
\end{equation}
We notice that the integrand in the integral
\[
\tilde{f}_q(-i\tau)=\int_{-a}^{a}\frac{f(x)dx}{[1-(q-1)\tau x
f^{q-1}(x)]^{\frac{1}{q-1}}}
\]
is strictly grater than $f(x)$ if $\tau>0,$ implying
$|\tilde{f}_q(-i\tau)|>1,$ since $f$ is a density function.
Therefore, the right hand side of \eqref{suppsize} is positive and
gives indeed an estimate of the size of the support of $f$ from
below.

Let $f_N(x)=f_{S_N}(x)$ be the density function of
$S_N=X_1+\dots+X_N,$ where $X_1,\dots,X_N$ are $q$-independent
random variables with the same density function $f=f_{X_1}$ whose
support is $[-a,a].$ We show that the $q$-independence condition can
not reduce the support of $f_N$ to an interval independent of $N.$
More precisely, $d(f_N)$ increases at the rate of $N$ when $N \to
\infty.$

\begin{thm}\label{Prop:PW1}
Let  $X_1,\dots,X_N$ {be} $q$-independent of any type I-III random
variables all having the same density function $f$ with $supp \, f
\subseteq [-a,a].$ Then, for the size of the density $f_N$ of $S_N$,
there exists a constant $K_q >0$ such that the estimate
\begin{equation} \label{prop1:estfN}
d(f_N) \ge K_q N  \sup_{\tau} \frac{\ln_q
|\tilde{f}_q(-i\tau)|}{\tau}
\end{equation}
holds.
\end{thm}

\textit{Proof.} Using formula \eqref{suppsize} one has
\begin{equation} \label{suppsize10}
d(f_N) \ge \frac{2}{M_{q,N}}\sup_{\tau} \frac{\ln_q
|\widetilde{({f}_{N})}_{q}(-i\tau)|}{\tau},
\end{equation}
where $M_{q,N}=\max_{x\in[-Na,Na]} f_N^{q-1}(x).$ It is clear from
probabilistic arguments that $M_{q,N} \le M_q$ for all $N\ge 2.$
Therefore, it follows from \eqref{suppsize10} that
\begin{equation} \label{suppsize15}
d(f_N) \ge \frac{2}{M_{q}}\sup_{\tau} \frac{\ln_q
|\widetilde{({f}_{N})}_{q}(-i\tau)|}{\tau},
\end{equation}
Let $X_N$ be $q$-independent of type I (see \eqref{q-ind1}). Making
use of the inequality $|z-r|\ge |z|-r,$ which holds true for any
complex $z$ and positive integer number $r,$ one has
\begin{align*}
|\widetilde{({f}_{N})}_{q}(-i\tau)|&=|\tilde{f}_q(-i\tau) \otimes_q
\dots \otimes_q \tilde{f}_q (-i \tau)|\\&=|[N
\left(\tilde{f}_q(-i\tau)\right)^{1-q}-(N-1)]^{\frac{1}{1-q}}|
\\&\ge [N
|\tilde{f}_q(-i\tau)|^{1-q}-(N-1)]^{\frac{1}{1-q}}\\&=|\tilde{f}_q(-i\tau)|
\otimes_q \dots \otimes_q |\tilde{f}_q (-i \tau)|.
\end{align*}
Taking $q$-logarith of both sides in this inequality and using the
property $\ln_q(g \otimes_q h)=\ln_q g + \ln_q h,$ one obtains
\begin{equation}
\label{suppsize16} \ln_q |\widetilde{({f}_{N})}_{q}(-i\tau)| \ge N
\ln_q |\tilde{f}_q(-i\tau)|.
\end{equation}
Now estimate \eqref{prop1:estfN} follows from inequalities
\eqref{suppsize15} and \eqref{suppsize16}.

For random variables $X_N$ independent of type II, equation
\eqref{suppsize16} takes the form
\begin{equation}
\label{suppsize20} \ln_{q_{_1}} |\widetilde{({f}_{N})}_{q}(-i\tau)|
\ge N \ln_{q_{_1}} |\tilde{f}_q(-i\tau)|.
\end{equation}
Since $1<q<q_1$ and $\frac{q-1}{q_1-1}=\frac{3-q}{2},$ making use of
inequalities \eqref{ln01} and \eqref{ln02} and assuming that
$|\widetilde{({f}_{N})}_{q}(i\tau)| \ge 1,$ one has
\begin{equation}
\label{suppsize21} \ln_{q} |\widetilde{({f}_{N})}_{q}(-i\tau)| \ge
\frac{(3-q)N}{2} \ln_{q} |\tilde{f}_q(-i\tau)|.
\end{equation}
Similarly, for variables independent of type III, we have
\begin{equation}
\label{suppsize30} \ln_{q} |\widetilde{({f}_{N})}_{q}(-i\tau)| \ge N
\ln_{q_{_1}} |\tilde{f}_{q_{_1}}(-i\tau)|.
\end{equation}
Both \eqref{suppsize21} and \eqref{suppsize30} obviously imply
estimate \eqref{prop1:estfN}. \eproof

\begin{cor}
\label{cor1} Let  $X_1,\dots,X_N$ be $q$-independent of any type
I-III random variables all having the same density function $f$ with
$supp \, f \subseteq [-a,a].$ If the sequence $Z_N$ has a
distributional limit random variable in some sense, then this random
variable can not have a density with compact support. Moreover, due
to the scaling present in $Z_N,$ the support of the limit variable
is the entire axis.
\end{cor}
The proof obviously follows immediately from \eqref{prop1:estfN}
upon letting $N \to \infty.$

\section{On Hilhorst's counterexamples}

We want to compare Theorem \ref{Prop:PW1} with Hilhorst's
counterexamples in \cite{Hilhorst2010}. He used the invariance
principle to show that $q$-FT is not invertible. Let $f(x), \, x \in
(-\infty,\infty),$ be a symmetric density function, such that
$\lambda(x)=x[f(x)]^{q-1}$ restricted to the semiaxis $[0,\infty)$
has a unique (local) maximum $m$ at a point $x_m.$
In other words $\lambda(x)$ has two monotonic pieces,
$\lambda_{-}(x), \, 0 \le x \le x_m,$ and $\lambda_{+}(x), \, x_m
\le x < \infty.$ Let $x_{\pm}(\xi), \, 0 \le \xi \le m,$ denote the
inverses of $\lambda_{\pm}(x),$ respectively. Then the $q$-FT
($1<q<2$) of $f$ can be expressed in the form, see
\cite{Hilhorst2010}
\begin{equation*}
\label{qFT1}
\tilde{f}_q(\xi)=\int_{-\infty}^{\infty}F(\xi')\exp(i\xi\xi')d\xi',
\end{equation*}
{where}
\begin{equation}
F(\xi)=\frac{q-2}{q-1}\xi^{\frac{1}{q-1}}\frac{d}{d\xi}[x_{-}^{\frac{q-1}{q-2}}(\xi)-
x_{+}^{\frac{q-1}{q-2}}(\xi)], \, \xi \in [0,m].
\end{equation}
The invariance principle yields
\begin{equation}
\label{qFT2}
F(\xi)=\frac{q-2}{q-1}\xi^{\frac{1}{q-1}}\frac{d}{d\xi}[{X}_{-}^{\frac{q-1}{q-2}}(\xi)-
X_{+}^{\frac{q-1}{q-2}}(\xi)], \, \xi \in [0,m],
\end{equation}
where
\begin{equation} \label{Xpm}
X_{\pm}^{\frac{q-1}{q-2}}(\xi)=x_{\pm}^{\frac{q-1}{q-2}}(\xi)+H(\xi),
\end{equation}
with $H(\xi)$ being a function defined on $[0,m],$ and such
that $X_{\pm}^{\frac{q-1}{q-2}}(\xi)$ are invertible. Denote by
$\Lambda (x)$ the function defined by the two pieces of inverses of
$X_{\pm}(\xi),$ namely
\begin{equation*}
\Lambda_{H} (x)=
\begin{cases}
X_{-}^{-1}(x),            &\text{if $0 \leq x \le x_{m,H}$,}\\
X_{+}^{-1}(x),            &\text{if $x > x_{m,H}$,}
\end{cases}
\end{equation*} 
where
$x_{m,H}=[(q-1)^{\frac{q-1}{2(2-q)}}+H(m)]^{-\frac{2-q}{q-1}}.$ The
function $\Lambda_H(x)$ is continuous, since
$X_{-}^{-1}(x_{m,H})=X_{+}^{-1}(x_{m,H}).$ Then
\begin{equation}
\label{fH} f_H(x)=\left(\frac{\Lambda(x)}{x} \right)^{\frac{1}{q-1}}
\end{equation}
defines a density function with the same $q$-FT as of $f.$ The
density $f_H$ coincides with $f$ if $H(\xi)$ is identically zero.

Now assume that $f(x)$ is a $q$-Gaussian,
$$f(x)=G_q(x)=\frac{C_q^{q-1}}{[1+(q-1)x^2]^{\frac{1}{q-1}}}, \, 1<q<2,$$
where $C_q$ is the normalization constant. Obviously, $G_q(x)$ is
symmetric, and the function $\lambda_q(x)=x[G_q(x)]^{q-1}$
considered on the semiaxis $[0,\infty)$ has a unique maximum
$m=\frac{C_q^{q-1}}{2\sqrt{q-1}}$ attained at the point
$x_m=(q-1)^{-\frac{1}{2}}$. Moreover, the functions $x_{\pm}(\xi)$
in this case take the forms (see \cite{Hilhorst2010})
\begin{equation}
\label{pieces1} x_{\pm}(\xi)=\frac{C_q^{q-1}\pm
[C_q^{2(q-1)}-4(q-1)\xi^2]^{1 \over 2}}{2\xi(q-1)}, \, 0 < \xi \le
m.
\end{equation}
We denote the density $f_H(x)$  and the function $\Lambda_{H}(x)$
corresponding to the $q$-Gaussian by $G_{q,H}(x)$ and
$\Lambda_{q,H}(x),$ respectively.  Hilhorst, selecting $H(\xi)=A \ge
0$ constant, constructed a family of densities
\begin{equation}
\label{fA} G_{q, A}(x)=\frac{C_q
(x^{\frac{q-2}{q-1}}-A)^{\frac{1}{q-2}}}{x^{\frac{1}{q-1}}[1+(q-1)(x^{\frac{q-2}{q-1}}-A)^{2\frac{q-1}{q-2}}]^{\frac{1}{q-1}}},
\end{equation}
which
have the same $q$-FT as the $q$-Gaussian for all $A.$ The following
statement shows that none of the densities $G_{q, A}(x)$ can be a
limit distribution in the $q$-CLT, except the one, corresponding to
$A=0,$ which coincides with the $q$-Gaussian, $G_{q,0}(x)=G_q(x)$.

\begin{prop}
\label{prop:comsup} Let $H(0) > 0.$ Then the support of $G_{q,
H}(x)$ is compact, and $$supp \, G_{q,H}=
\left[-[H(0)]^{\frac{q-1}{q-2}},[H(0)]^{\frac{q-1}{q-2}}\right].
$$
\end{prop}

\textit{Proof.} Since $\lim_{\xi \to 0} x_{+}(\xi)=+\infty,$ the
largest value of $X_{+}$ is equal to $\lim_{\xi \to
0}X_{+}(\xi)=[H(0)]^{\frac{q-1}{q-2}}.$ Therefore, the inverse of
$X_{+}$ is defined on the interval $[x_0,
[H(0)]^{\frac{q-1}{q-2}}],$ where $x_0>0$ is some number obtained by
{\bf a} shifting of $x_m$ depending on $H(m).$ On the other hand the
smallest value of $x_{-}$ is zero, taken at $\xi=0.$ Therefore, the
inverse of $X_{-}$ is defined on the interval $[0,x_0].$ Hence,
 by symmetry, $G_{q,H}$ has the support
$\left[-[H(0)]^{\frac{q-1}{q-2}},[H(0)]^{\frac{q-1}{q-2}}\right].$
\eproof
\begin{remark}
{Note that $H(0)$ can not be negative. In fact, if $H(0)<0,$ then
{either} $X_{\pm}$ is not invertible or, if it is invertible, its
inverse does not identify a density function.}
\end{remark}
{Proposition \ref{prop:comsup} implies that if $H(0) > 0$ then, due
to Corollary \ref{cor1}, $G_{q,H}(x)$ can not be the density
function of the limit distribution in the $q$-CLT. Thus none of the
densities in Hilhorst's counterexamples\footnote{See Examples 2 and
3 in \cite{Hilhorst2010}. Example 4 is not relevant to the $q$-CLT,
since in this case, $(2q-1)$-variance of the 2-Gaussian does not
exist, and consequently the $q$-CLT is not {applicable}.} except the
$q$-Gaussian can serve as an
attractor
in the $q$-CLT.}

Only one possibility is left, {namely} $H(0)=0.$ The next
proposition establishes that, in this case, $G_{q,H}(x)$ is
asymptotically equivalent to $G_q(x) \equiv G_{q,0}(x).$

\begin{prop} \label{prop:asequiv}
Let $H(0)=0.$ Then
$$\lim_{|x|\to \infty} \frac{G_{q, H}(x)}{G_q(x)}=1. $$
\end{prop}

\textit{Proof.} {Since} $H(0)=0$, then obviously
\begin{equation*}
\lim_{\xi \to 0} \frac{X_{+}(\xi)}{x_{+}(\xi)}=\lim_{\xi \to 0}
\left(1+\frac{H(\xi)}{x_{+}(\xi)} \right)=1.
\end{equation*}
Therefore, for inverses one has
\begin{equation*}
\lim_{x \to +\infty} \frac{X_{+}^{-1}(x)}{x_{+}^{-1}(x)}=1.
\end{equation*}
This implies
\begin{equation*}
\lim_{x \to +\infty} \frac{G_{q,H}(x)}{G_q(x)}= \lim_{x \to +\infty}
\left(\frac{\frac{X_{+}^{-1}(x)}{x}}{\frac{x_{+}^{-1}(x)}{x}}\right)^{\frac{1}{q-1}}=1.
~~~~~~~~~~~~~~~~  \eproof
\end{equation*}

\begin{remark}

Propositions \ref{prop:comsup} and \ref{prop:asequiv} establish that
$G_{q,H}$ can identify a limiting distribution in the $q$-CLT only
if $H(0)=0.$ However, in this case, independently {from} other
values of $H(\xi),$ the density $G_{q,H}(x)$ is asymptotically
equivalent to the $q$-Gaussian, i.e. $G_{q,H}(x) \sim G_{q}(x) \, \,
\mbox{as} \, \, |x|\to\infty.$ We return to this question in the
Conclusion, where we discuss whether $G_{q,H}$ can, for $H$ not
identically zero, be an attractor in the $q$-CLT.

\end{remark}

The statement of the following proposition can be proved exactly as
Proposition \ref{prop:asequiv}, replacing $X_+(\xi), \, x_+(\xi)$ by
functions $X_-(\xi), \, x_-(\xi),$ respectively.

\begin{prop} \label{prop:asequivat0}
Let $H(0)=0.$ Then
$$\lim_{x\to 0} \frac{G_{q, H}(x)}{G_q(x)}=1. $$
\end{prop}

\section{Other relations of $G_{q,H}$ {\bf with} the $q$-Gaussian}

In light of Propositions \ref{prop:comsup} and \ref{prop:asequiv},
we will assume {below} that $H(0)=0$ and clarify other conditions
for $H(\xi).$ As above we use notations
$\Lambda_{q,H}(x)=x[G_{q,H}(x)]^{{q-1}}$ and
$\lambda_q(x)=x[G_q(x)]^{q-1}.$

\begin{prop}
Let $H(m)=0.$ Then the function $\Lambda_{q,H}(x)$ attains its
unique maximum at the point $x_m,$ and
$\Lambda_{q,H}(x_m)=\lambda_q(x_m)=m.$

\end{prop}

\textit{Proof.} If $H(m)=0,$ then it follows from \eqref{Xpm}
immediately, that $X_{\pm}(m)=x_{\pm}(m),$ which implies
$\Lambda_{q,H}(x_m)=\lambda_q(x_m)=m.$ \eproof

\medskip

The statement below clarifies the range of {the} values $H(\xi)$.
\begin{prop} \label{Hval}
The function
$$f_{q,H}(x)=\Big(\frac{\Lambda_{q,H}(x)}{x}\Big)^{\frac{1}{q-1}}$$
defines a density function if $H(\xi)$ for all $\xi \in (0,m]$
satisfies the following condition
\begin{equation} \label{values}
-\frac{1}{[x_+(\xi)]^{\frac{q-1}{2-q}}} < H(\xi) <
\left(\frac{C_q^{q-1}}{\xi}\right)^{\frac{q-1}{2-q}}-\frac{1}{[x_-(\xi)]^{\frac{q-1}{2-q}}}.
\end{equation}
\end{prop}

\textit{Proof.} Let $0<\xi\le m.$ The conditions $X_{-}(\xi)
>C_q^{1-q} \xi$ and $X_{+}(\xi)<\infty$ together with \eqref{Xpm} imply estimate \eqref{values}.
\eproof

\begin{remark} Proposition \ref{Hval} implies that the range of
$H(m)$ is restricted to the interval
\[
-(q-1)^{\frac{q-1}{2(2-q)}} < H(m) < (q-1)^{\frac{q-1}{2(2-q)}}.
\]

\end{remark}

\begin{prop}
Let $H(m)\ne 0.$ Then the function $\Lambda_{q,H}(x)$ attains its
unique maximum at the point
$x_{m,H}=\Big[(q-1)^{\frac{q-1}{2(2-q)}}+H(m)\Big]^{-\frac{2-q}{q-1}},$
and $\Lambda_{q,H}(x_{m,H})=\lambda_q(x_m)=m.$
\end{prop}

\textit{Proof.} Since $x_{m,H}=X_{-}(m)=X_{+}(m),$ the statement of
this proposition can easily be derived upon computing $X_{-}(m).$
\eproof

\begin{prop} \label{prop:HisPositive}
Let $q\in(3/2,2).$ The inequality $X_{+}^{'}(\xi) <0$ holds near
zero if and only if $H \ge 0$ near zero.
\end{prop}

\textit{Proof.} It is not hard to verify that
$x_{+}(\xi)=\frac{A_q}{\xi}+O(\xi)$ and
$x_{+}^{'}(\xi)=-\frac{A_q}{\xi^2}+O(1),$ as $\xi \to 0,$ where
$A_q=C_q^{q-1}/(q-1).$ Differentiating both sides of equation
\eqref{Xpm} for $X_{+},$ one has
\begin{equation} \label{conc10}
X_{+}^{'}(\xi)=  \left( \frac{X_{+}(\xi)}{x_{+}(\xi)}
\right)^{\frac{1}{2-q}}x_{+}^{'}(\xi)- H^{'}(\xi)
\frac{(2-q)\left(X_{+}(\xi)\right)^{\frac{1}{2-q}}}{q-1}.
\end{equation}
Due to Proposition \eqref{prop:asequivat0} $X_{+}(\xi) \sim
x_{+}(\xi)$ as $\xi \to 0.$ Therefore,
\begin{align}
X_{+}^{'}(\xi) &\sim  x_{+}^{'}(\xi)- H^{'}(\xi)
\frac{(2-q)\left(x_{+}(\xi)\right)^{\frac{1}{2-q}}}{q-1} \notag \\
&\sim -\frac{A_q}{\xi^2}- H^{'}(\xi)
\frac{B_q}{\xi^{\frac{1}{2-q}}}, ~~ \xi \to 0,   \label{conc11}
\end{align}
where $B_q=\frac{(2-q)A_q^{\frac{1}{2-q}}}{q-1}.$ If $H(\xi)\ge 0,$
then obviously, $X_{+}^{'}(\xi) < 0$ near zero. Now assume that
$H(\xi)<0$ near zero (that is in an interval $(0,\varepsilon)$ with
some $\varepsilon >0$). Then the second term in \eqref{conc11} grows
faster then the first term near zero if $q>3/2,$ implying
$X_{+}^{'}(\xi)\ge 0$ near zero. \eproof

\begin{prop} \label{Hyperbolic}
Let $x_1^a$ and $x_2^a$ be two numbers such that
$0<x_1^a<x_{m,H}<x_2^a,$ and
\[
\Lambda_{q,H}(x_1^a)=\Lambda_{q,H}(x_2^a)=a,
\]
then $x_1^a x_2^a=const$ for all values of $a \in(0,m]$ if and only
if $H(\xi)$ is identically zero.
\end{prop}

\textit{Proof.} Let $H(\xi) \equiv 0.$ Then
$$\Lambda_{q,H}(x)=\lambda_q(x)=
\frac{C_q^{q-1}x}{1+(q-1)x^2}.$$ In this case the conclusion of the
proposition can be established by direct calculation. Indeed, $x_1^a
x_2^a=(q-1)^{-1},$ which is independent of the values of $a.$ Now
assume that $H(\xi) \ne 0.$ Then, using \eqref{Xpm} it is readily
seen that
\begin{align*}
X_+(a)X_-(a)=\Big([x_+(a)x_-(a)]^{\frac{q-1}{q-2}}+\mu(a)\Big)^{\frac{q-2}{q-1}}
\end{align*}
where $\mu(a)=H(a)\left(
x_+^{\frac{q-1}{q-2}}(a)+x_+^{\frac{q-1}{q-2}}(a) + H(a)\right).$
Since $x_-(a)$ and $x_+(a)$ equal respectively $x_1^a$ and $x_2^a$
corresponding to the case $H(\xi) \equiv 0,$ it follows that
\begin{equation} \label{hyper1}
X_+(a)X_-(a)=(const + \mu(a))^{\frac{q-2}{q-1}}.
\end{equation}
Now inverting $X_{\pm}(x)$ we obtain that the product $x_1^a x_2^a$
is dependent of $a.$

Equation \eqref{hyper1} also implies the necessity of the condition
$H(\xi) \equiv 0$ for $x_1^a x_2^a$ to be independent on $a.$
\eproof

\section{Conclusion}

Concluding, we would like to note some key points related to the
limiting distribution in the $q$-CLT, the role of the $q$-FT in this
as well as other relevant theorems, and also briefly address other
concerns raised by Hilhorst in his paper \cite{Hilhorst2010}.

\begin{enumerate}
\item
In the proof of the $q$-CLT (see \cite{UmarovTsallisSteinberg2008}),
the $q$-FT is used only to establish the existence of a limiting
distribution. If we assume that there is another (non unique)
limiting distribution, then, due to Propositions \ref{prop:asequiv}
and \ref{prop:asequivat0}, this distribution shares the same value
at the origin and the same asymptotic behavior at infinity as the
$q$-Gaussian. Therefore, such a density {may be seen as} a {
(slight)} deformation of the $q$-Gaussian.

However, can a distribution defined by $G_{q,H},$ if $H$ is not
identically zero, be a limit distribution in the $q$-CLT? Our belief
is that this can not happen.
Two strong arguments in favor of this belief are the following ones:
\begin{enumerate}
\item[(i)]
Proposition \ref{prop:HisPositive} states that $H(\xi)$ can not be
negative near the origin for $3/2<q<2$. Due to the {plausible}
nature of attractors, it is very unlikely that $H(\xi)$ has
non-smooth points and drastic changes including sign changes. If
$H(\xi)$ is not smooth at some points then $G_{q,H}$ will have
singular points. A change of sign adds a new inflection point in the
graph of the density of a limiting distribution. Therefore, if $H$
is not negative in some small interval then it is not negative on
the whole interval $[0,m].$
All these facts essentially restrict the set of
functions $H(\xi)$ used in the invariance principle (possibly
reducing this set to $\{0\}$), for which $G_{q,H}$ would be an
attractor.

\item[(ii)]
Due to the strong dependence between $q$-independent random
variables, a possible {\it sym\-met\-ry-like} relationship between
probabilities {for} small values and {those} for large values of the
scaled sums $Z_N$ is expected. Since the function
$\lambda_q(x)=x[G_q(x)]^{{q-1}}$ is used extensively, this function
may possess such a "symmetry", {if} it exists. In fact if $x_1^a$
and $x_2^a$ are two numbers such that $0<x_1^a<x_m<x_2^a,$ and $
\lambda_q(x_1^a)=\lambda_q(x_2^a)=a, $ then the product $x_1^a
x_2^a=\frac{1}{q-1}$ is constant for all values of $a \in(0,m].$ We
call this property  \textit{the hyperbolicity property} (since
$x_2^a=\frac{1}{(q-1)x_2^a}$ is a hyperbola in $(x_1^a,
x_2^a)$-plane).
This heuristic argument leads to the following conjecture:
\smallskip

\textit{Conjecture: The limit distribution of the sequence $Z_N$ in
equation \eqref{ZN} for $q$-independent random variables $X_N$
possesses the hyperbolicity property.}
\medskip

\noindent Proposition \ref{Hyperbolic} shows that the only density
with this property in the class of functions $G_{q,H}, \, H(0)=0,$
is the $q$-Gaussian. If the above conjecture is true, then the
uniqueness of a limit distribution in the $q$-CLT will immediately
follow from Proposition \ref{Hyperbolic}.
\end{enumerate}

\item
Regarding the paper \cite{UmarovTsallisGellMannSteinberg2010}, we
recall that it is devoted to the asymptotical analysis of limiting
distributions of $(q,\alpha)$-{stable} distributions, explicitly
{addressing}
the "asymptotic equivalence." This paper, like 
\cite{UmarovTsallisSteinberg2008}, 
essentially uses the $q$-FT technique to establish the existence of
a limiting distribution. To this end, we recall that {even} the
usual FT (i.e., the $1$-FT) can {not} be used for the rigorous proof
of the uniqueness of solution in many situations, {even though it is
invertible}. Therefore, in the classical theory the Fourier series
or Fourier transform techniques are used for the existence of a
solution. The use of this technique for the uniqueness is restricted
to classes of functions which are representable as a Fourier series
or Fourier transformable. For the uniqueness usually other methods
are involved, like the "maximum principle", "energy integral", etc.
Concluding this remark, we would like to note that  in the
references
\cite{UmarovTsallis2007,UmarovTsallisGellMannSteinberg2010,UmarovQueiros2010}
mentioned by Hilhorst in his paper \cite{Hilhorst2010}, the $q$-FT
is used only for existence purposes.

\item
Another question raised by Hilhorst in his papers
\cite{Hilhorst2010,Hilhorst2009} is the lack of examples of
$q$-independent random variables and their applications.
Mathematically, a non-vacuous definition of some notion is a recipe
for producing examples.
The notion of $q$-independence generalizes the notion of usual
independence, hence, containing as a trivial particular case the
usual independence. Non-trivial examples can be produced at will
using the definition of $q$-independence. As an example, for
practical applications, paper \cite{VignatPlastino2007} proves that
sequences of independent random variables mixed with the help of a
chi-square distribution are asymptotically $q$-independent. Paper
\cite{HahnJiangUmarov2010} shows that such sequences can be
considered as variance mixtures of normals, a wide class of
distributions with applications in Beck-Cohen superstatistics
\cite{BeckCohen}. It is not surprising that sequences of variance
mixtures of normals have limit distributions with $q$-Gaussian
densities (see \cite{HahnJiangUmarov2010}).

\item
Last but not least. It is definitively clear that no
experimental, observational or computational results will ever
determine an analytical function unless we have strong (physical)
reasons to
severely restrict its class. 
Independently from the $q$-CLT, {a wide spectrum of experimental and
computational distributions have been interpreted in the literature
as $q$-Gaussians}. Hence, various strong { analytical} reasons do
exist which make $q$-Gaussians quite special. These include:
\begin{enumerate}
\item[(i)]
Under simple width constraints, $q$-Gaussians extremize the
nonadditive entropy $S_q$, whose uniqueness (under natural axioms)
and physical relevance has been repeatedly exhibited in the
literature from many standpoints (see
\cite{GellMannTsallis2004,CarusoTsallis2008,SaguiaSarandy2010,Tsallis2009a,Tsallis2009b});
\item[(ii)] $q$-Gaussians have been
shown to exactly solve, for all values of space and time, the
so-called {\it Porous Medium Equation} \cite{PlastinoPlastino1995,TsallisBukman1996}, a very basic nonlinear
Fokker-Planck equation (which satisfies the $H$-theorem precisely for
the entropy $S_q$~\cite{SchwammleNobreCurado2007,SchwammleCuradoNobre2009}, and which can be deduced from a quite simple
non-Markovian Langevin equation)
\cite{FuentesCaceres2008};

\item[(iii)] Scale-invariant
probabilistic models have been analytically shown to yield
$q$-Gaussian limiting distributions for large systems, in a way
totally analogous to how the de Moivre-Laplace theorem yields
Gaussians, with the latter result being recovered as the $q=1$ case
of these models
\cite{RodriguezSchwammleTsallis2008,HanelThurnerTsallis2009};
\item[(iv)] $q$-Gaussians consistently are attractors of $q$-CLT´s that do not use the $q$-FT in their proofs \cite{VignatPlastino2007,HahnJiangUmarov2010}.
\end{enumerate}
\end{enumerate}

We acknowledge worthful remarks from M. Hahn, X. Jiang, and K.
Nelson.

\end{document}